# Reconstructing Minkowski geometry from causal separations

*Chenyang (Amy) Hu*[1,2], *David A. Meyer*[3] and *Eleanor J. Q. Meyer*[4]

[1]Westview High School, San Diego, CA 92129
[2]Carnegie Mellon University, Pittsburgh, PA 15213
[3]Department of Mathematics, UC San Diego, La Jolla, CA 92093-0112
[4]The Bishop's School, La Jolla, CA 92037

`amyh2@andrew.cmu.edu`, `dmeyer@ucsd.edu`, `eleanor.meyer.26@bishops.com`

Aleksandrov, and then Zeeman, showed that the causal relations among the set of points in a Minkowski space of dimension greater than 2 determine the Minkowski space structure of the set up to a global conformal factor. We show that in any dimension the distances between causally related pairs of points determine the distances between spatially related pairs of points, and thus completely determine the Minkowski space structure of the set. This is a step in the direction of proving that causal sets arising from a Poisson process in a Lorentzian manifold determine that manifold up to the degree of approximation inherent in the intensity of the Poisson process—the *Hauptvermutung* of causal set theory.

## 1. Introduction

In the causal set approach to a quantum theory of gravity the underlying structure of spacetime is a discrete partially ordered set (poset) where the partial order relation, $\preccurlyeq$, represents 'causally precedes' and the number of elements in a subset counts spacetime volume in Planck units [1]. The goal is a quantum sum-over-histories formulation in which, at scales much above the Planck scale, interference reënforces amplitudes of posets that are well-approximated by a Lorentzian manifold satisfying the Einstein equation of general relativity. 'Well-approximated' should mean that these posets are close, in some metric, to ones that are likely to arise as the set of points in a density 1 Poisson process in the manifold, partially ordered by causality in the manifold. The *Hauptvermutung* (central conjecture) of causal set theory is that the manifold is unique up to this degree of approximation [2].

Belief in the *Hauptvermutung*[1] rests first on the theorem of Hawking [6] and Malament [7] that in Lorentzian manifolds of dimension at least 3 the causal structure alone determines the metric up to a local conformal factor, and second on the idea that identifying the number of elements in a subset of a causal set with the spacetime volume should determine that factor [1]. Our goal here is to prove a novel, albeit modest, theorem supporting this belief.

---

[1] Despite the disproof [3] of its namesake in geometric topology [4,5]!

We take inspiration from two results, one old and one new. The first is Aleksandrov's theorem, announced in 1950 [8], proved with Ovchinnikova in 1953 [9], and rediscovered by Zeeman in 1964 [10], on which the results of Hawking and Malament are based, that in Minkowski spaces of dimension at least 3 the causal structure determines the space up to homothecy, *i.e.*, up to a dilation (global conformal factor). The second is a new method for measuring distances between causally unrelated (*i.e.*, spacelike separated) elements in a causal set, proposed recently by Boguñá and Krioukov, which begins by estimating distances between causally related elements and assumes an underlying Minkowskian geometry [11]. That is, we focus on Minkowski spaces, $\mathbb{M}^d$, and show that their causal structure, together with the distances between points that are causally related, determines the distances between spacelike separated points and thus the global conformal factor, for any $d \in \mathbb{N}$.

The estimate of the distance between two causally related elements used by Boguñá and Krioukov [11] is proportional to one less than the number of elements in a longest chain of elements connecting them, where the proportionality constant depends upon the density of the Poisson process in a Minkowski space of which the causal set is presumed to be an outcome, and on the dimension of that space [12,13]. That is, there is a formula combining the combinatorics of the causal set and the dimension of the originating Minkowski space to give an estimate of causal distance. Given causal distances in a Minkowski space, but without the combinatorial information in a causal set generated by a Poisson process, this formula cannot be inverted to determine the dimension of the Minkowski space. This is important, because the proofs of Aleksandrov's [9], Zeeman's [10], and Hawking's [6] theorems take the dimension of the spacetime as given, so we should expect that knowing the dimension is necessary for us to use the causal distances to determine the global conformal factor. In §3 we show that the dimension of a Minkowski space can be determined from its set of points with their causal relations, and their causal (but not spacelike) distances. Then, having determined the dimension, we show in §4 how to compute the distances between spatially related points, and determine the vector space structure of the space. To motivate our approach, we begin in §2 by analyzing the case of $\mathbb{M}^2$. Not only does this set the stage for the general analysis, but it also demonstrates that our results apply in dimensions 1 and 2, unlike Aleksandrov's [8,9], Zeeman's [10], Hawking's [6] and Malament's [7].

## 2. Preliminaries

For notational convenience we identify $d$-dimensional Minkowski space, $\mathbb{M}^d$, with its tangent space, a $d$-dimensional real vector space. The *Minkowski inner product* is a non-degenerate bilinear form $\eta$ with signature $(-+\cdots+)$ defining a scalar product on $\mathbb{M}^d$, $\eta : \mathbb{M}^d \times \mathbb{M}^d \to \mathbb{R}$. A vector $v \in \mathbb{M}^d$ is *timelike* if $\eta(v,v) < 0$, *spacelike* if $\eta(v,v) > 0$, and *null* or *lightlike* if $\eta(v,v) = 0$; it is *causal* if it is timelike or null. The *spacetime separation* of $p, q \in \mathbb{M}^d$ is $\eta(q-p, q-p)$. If $p \in \mathbb{M}^d$, then the *causal future* of $p$ is

$$J^+(p) = \{q \in \mathbb{M}^d \mid q - p \text{ is causal and } q_0 \geq p_0\},$$

where the vector index runs from 0 to $d-1$. The *causal past* of $p$, $J^-(p)$, is defined analogously. Similarly, the *timelike future* of $p$ is

$$I^+(p) = \{q \in \mathbb{M}^d \mid q - p \text{ is timelike and } q_0 \geq p_0\},$$

and the *timelike past*, $I^-(p)$, is again defined analogously. When $q \in J^+(p)$ we write $p \preccurlyeq q$ (when $q \in I^+(p)$ we write $p \prec q$). Thus $\eta$ determines the *causal structure* of $\mathbb{M}^d$, the relation $\preccurlyeq$.

We begin by noting that if its causal structure and causal distances are given, then $\mathbb{M}^1$ is completely determined: if we pick an arbitrary point, $o$, and assign it coordinate $(0)$, then any other point $p$ has some timelike separation $-s^2$, $0 < s \in \mathbb{R}$, from $o$, and we assign it coordinate $(t_p) = (s)$ if it lies to the future of $o$ or $(t_p) = (-s)$ if it lies to the past. This defines the vector space structure of $\mathbb{M}^1$: $p + q$ has coordinate $(t_p + t_q)$.

Thus the lowest nontrivial dimension is 2. Consider two spacelike related points $a, b \in \mathbb{M}^2$. Our goal is to compute the spacelike distance between them, using the causal structure of $\mathbb{M}^2$, and distances between causally related points.

The spacelike distance between $a$ and $b$ can be computed using the properties of hyperbolae in Minkowski space. The following lemma establishes an elementary, but important, property of these hyperbolae. For $p \in \mathbb{M}^d$, define $H_p(-s^2)$ to be the hyperbola of points that are timelike separated by $-s^2 < 0$ from $p$, and let $H_p^\pm(-s^2) = H_p(-s^2) \cap I^\pm(p)$ be the past and future sheets of this hyperbola. Since each sheet is spacelike, we call $H_p(-s^2)$ a *spacelike* hyperbola.

SPACELIKE HYPERBOLAE LEMMA. *When $a$ and $b$ are spacelike related in $\mathbb{M}^d$, for any $0 < s_1, s_2 \in \mathbb{R}$, $H_a(-s_1^2)$ and $H_b(-s_2^2)$ intersect in a $(d-2)$-dimensional hyperboloid, $H_{ab}(s_1, s_2) = H_{ab}^-(s_1, s_2) \cup H_{ab}^+(s_1, s_2)$, where $-$ and $+$ denote the past and future sheets, respectively.*

*Proof.* Since $a$ and $b$ are spacelike related, we can choose a coordinate system where their coordinates are $(0, \ldots, 0, -h)$ and $(0, \ldots, 0, h)$ for some $h > 0$. Then the equations for these two hyperbolae are:

$$-t^2 + x_1^2 + \cdots + x_{d-2}^2 + (x_{d-1} - h)^2 = -s_1^2$$
$$-t^2 + x_1^2 + \cdots + x_{d-2}^2 + (x_{d-1} + h)^2 = -s_2^2.$$

Subtracting the first equation from the second gives

$$(x_{d-1} + h)^2 - (x_{d-1} - h)^2 = s_1^2 - s_2^2 \quad \Longrightarrow \quad 4h x_{d-1} = s_1^2 - s_2^2,$$

which has one solution for $x_{d-1}$. Substituting it back into either of the original equations gives the equation for a spacelike hyperboloid in the $x_{d-1} = (s_2^2 - s_1^2)/(4h)$ hyperplane:

$$-t^2 + x_1^2 + \cdots + x_{d-2}^2 = -\frac{(s_1^2 + s_2^2)^2 + 8h^2(s_1^2 + s_2^2 + 2h^2)}{16h^2}.$$

And we remark that each point of the past sheet of this spacelike hyperboloid is timelike related to every point of its future sheet since they are all timelike related to the center of the hyperboloid at $(0, \ldots, 0, (s_2^2 - s_1^2)/(4h))$. ∎

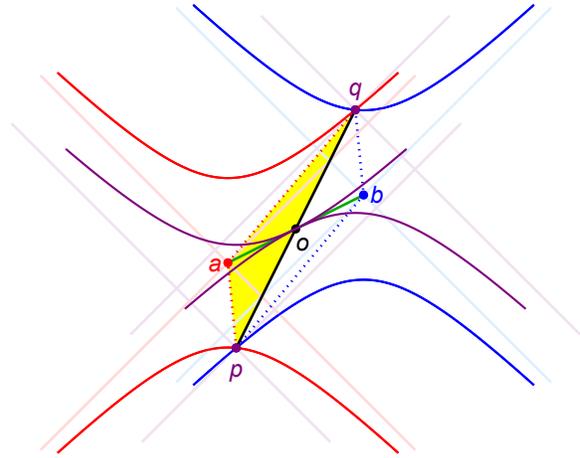

**Fig. 1.** Finding $\eta(b - a, b - a)$.

When $d = 2$, this lemma says that the hyperbolae $H_a(-s^2)$ and $H_b(-s^2)$ intersect in two points $p$ and $q$, to the past and future, respectively, of both $a$ and $b$, as shown in Figure 1. Since $p \prec q$, the separation $\eta(q - p, q - p) = -B^2 < 0$ is given. As in the proof of the lemma, there is a coordinate system in which $a$ and $b$ are at $(0, -h)$ and $(0, h)$, respectively, while $p$ and $q$ are at $(-B/2, 0)$ and $(B/2, 0)$, respectively. Thus $\mathsf{area}(pqa) = Bh/2$, where $2h$ is the (as yet unknown) spacelike distance between $a$ and $b$, so if we knew the area of triangle $pqa$, shown in yellow in Figure 1, we could solve for $2h$. But the square of this area is given by the Cayley-Menger determinant [14,15] in Minkowski space (which we review in the Appendix):

$$\left(\mathsf{area}(pqa)\right)^2 = \frac{1}{16} \det \begin{pmatrix} 0 & 1 & 1 & 1 \\ 1 & 0 & -s^2 & -B^2 \\ 1 & -s^2 & 0 & -s^2 \\ 1 & -B^2 & -s^2 & 0 \end{pmatrix} = \frac{1}{16} B^2 (B - 2s)(B + 2s),$$

where the last three rows and columns are labeled $p$, $a$, $q$, and the corresponding entries are their spacetime separations. This implies $\eta(b - a, b - a) = (2h)^2 = (B - 2s)(B + 2s)$.

This determines the spacelike distance between $a$ and $b$ in the case $d = 2$, but we can do more. To reconstruct the vector space structure we need only assign coordinates to points; we do so in the same coordinate system as in the proof of the lemma: $a = (0, -h)$, $p = (-B/2, 0)$, and $q = (B/2, 0)$. Now any other point, $r = (t, x)$, has spacetime separations from each of these points: $\ell_{ra}$, $\ell_{rp}$, and $\ell_{rq}$, which we are given when they are causal, and which we have just shown how to compute when they are spacelike. Then its coordinates must satisfy:

$$\ell_{rp} = -(t + B/2)^2 + x^2$$
$$\ell_{rq} = -(t - B/2)^2 + x^2$$
$$\ell_{ra} = -t^2 + (x + h)^2.$$

We can solve the first two equations for $t = (\ell_{rq} - \ell_{rp})/(2B)$. Substituting this into all three equations gives two quadratic equations for $x$, with a unique solution. Thus we can assign coordinates, uniquely, to any element of $\mathbb{M}^2$.

If we try to generalize this argument to $\mathbb{M}^3$, we run into a problem: Now, by the Spacelike Hyperbolae Lemma, the hyperboloids $H_a(-s^2)$ and $H_b(-s^2)$ (red and blue, respectively, in Figure 2) intersect in a 1-dimensional hyperboloid (a hyperbola) $H_{ab}(s,s)$ (bright green in Figure 2). We can pick one point on each sheet of $H_{ab}(s,s)$, for example, $p \in H_{ab}^-(s,s)$ and $q \in H_{ab}^+(s,s)$ (cyan in Figure 2); and since $p \prec q$ we know their timelike separation. As shown in Figure 2, $a$, $b$, $p$, and $q$ need not be coplanar, however, so this timelike separation, along with the timelike separations $-s^2$ between $a,b$ and $p,q$, does not determine the spacelike separation between $a$ and $b$. We need a third point in the same plane as $H_{ab}(s,s)$; it will form a triangle (yellow in Figure 2) with $p$ and $q$, and if we know its spacetime separations from all the other points, we can compute the volume of the congruent tetrahedra with this triangle as the base and either $a$ or $b$ as the fourth vertex, using the Cayley-Menger determinant. Every other point in the intersection $H_{ab}(s,s)$ is *spatially* related to either $p$ or $q$, *e.g.*, $r$ in Figure 2, however, whence that distance is unknown. Thus we need a different way of picking the third base point in this case, and in general, the $d$ base points necessary in $\mathbb{M}^d$; we find a way to do so in the next section.

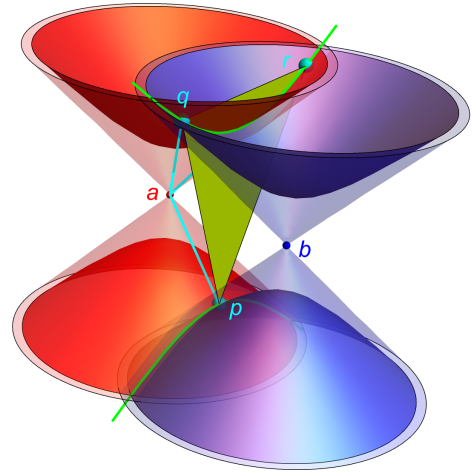

**Fig. 2.** The challenge in $d=3$.

## 3. Timelike gems

<div style="text-align:right">

然此帝網皆以寶成。
*However, this universal network is made of gems.*[2]

</div>

$H_{ab}(s,s)$ spans a timelike hyperplane, as illustrated in Figure 2; this is isometric to a Minkowski space of one lower dimension than the original space, so let us just investigate arrangements of timelike related points in $\mathbb{M}^d$ for any $0 < d \in \mathbb{N}$. We begin by choosing an arbitrary point, $p_0$. Next, choose a point $p_1 \in H_{p_0}^+(-1)$. Now consider $H_{p_0}^+(-s_2^2) \cap H_{p_1}^+(-1)$. There is a coordinate system in which $p_0 = (0,0,\ldots,0)$ and $p_1 = (1,0,\ldots,0)$, whence this intersection consists of points with coordinates solving:

$$\begin{aligned} -s_2^2 &= -t^2 + x_1^2 + \cdots + x_{d-1}^2 \\ -1 &= -(t-1)^2 + x_1^2 + \cdots + x_{d-1}^2. \end{aligned} \quad (3.1)$$

Subtracting the first equation from the second gives $s_2^2 = 2t$, and substituting $t = s_2^2/2$

---

[2] Our loose translation of the description of Indra's net attributed to 杜順 (Dùshùn, 557–640), first patriarch of Huáyán Buddhism [16].

into the first equation gives:

$$\frac{1}{4}s_2^4 - s_2^2 = \frac{1}{4}s_2^2(s_2^2 - 4) = x_1^2 + \cdots + x_{d-1}^2. \tag{3.2}$$

The right hand side of this equation is nonnegative, so we must have $s_2^2 \geq 4$ in order that $H_{p_0}^+(-s_2^2) \cap H_{p_1}^+(-1) \neq \emptyset$.

Alternatively, suppose there is a point $p_2 \in H_{p_0}^+(-s_2^2) \cap H_{p_1}^+(-1)$. Then the Cayley-Menger determinant giving the spacetime volume (area) squared of $p_0 p_1 p_2$ is

$$\bigl(\text{stvol}(p_0 p_1 p_2)\bigr)^2 = \frac{(-1)^2}{2^2(2!)^2} \det \begin{pmatrix} 0 & 1 & 1 & 1 \\ 1 & 0 & -1 & -s_2^2 \\ 1 & -1 & 0 & -1 \\ 1 & -s_2^2 & -1 & 0 \end{pmatrix} = \frac{1}{16} s_2^2(s_2^2 - 4),$$

which is proportional to the left hand side of (3.2), so to make this spacetime volume squared nonnegative, again we must have $s_2^2 \geq 4$.

Notice that if $d = 1$, the spacetime volume cannot be positive, so there is only a solution if $s_2^2 = 4$. So pick $s_2^2 = 6 > 4$.[3] If $H_{p_0}^+(-6) \cap H_{p_1}^+(-1) = \emptyset$, then $d = 1$. Otherwise $d > 1$ and we can pick a point, $p_2$, in the intersection.

Having found $p_2$, consider picking $p_3 \in H_{p_0}^+(-s_3^2) \cap H_{p_1}^+(-6) \cap H_{p_2}^+(-1)$. Then

$$\bigl(\text{stvol}(p_0 p_1 p_2 p_3)\bigr)^2 = \frac{(-1)^3}{2^3(3!)^2} \det \begin{pmatrix} 0 & 1 & 1 & 1 & 1 \\ 1 & 0 & -1 & -6 & -s_3^2 \\ 1 & -1 & 0 & -1 & -6 \\ 1 & -6 & -1 & 0 & -1 \\ 1 & -s_3^2 & -6 & -1 & 0 \end{pmatrix} = -\frac{1}{144}(s_3^2 - 25)(s_3^2 - 13).$$

For this to be nonnegative, we must have $13 \leq s_3^2 \leq 25$; choosing $s_3^2 = 19$ at the center of this range determines a third natural number in a sequence that gives positive Cayley-Menger determinants:

THEOREM 1. *Let* $g_i = 2^{i+2} - 3i - 4$, *for* $i \in \mathbb{N}$, *so that* $g_0 = 0$, $g_1 = 1$, $g_2 = 6$, $g_3 = 19$, ....[4] *Then the Cayley-Menger determinant*

$$\frac{(-1)^n}{2^n(n!)^2} \det \begin{pmatrix} 0 & 1 & 1 & 1 & \cdots & 1 \\ 1 & 0 & -g_1 & -g_2 & \cdots & -g_n \\ 1 & -g_1 & 0 & -g_1 & \ddots & \vdots \\ 1 & -g_2 & -g_1 & 0 & \ddots & -g_2 \\ \vdots & \vdots & \ddots & \ddots & \ddots & -g_1 \\ 1 & -g_n & \cdots & -g_2 & -g_1 & 0 \end{pmatrix} = \frac{3^{n-1}}{(n!)^2} > 0. \tag{3.3}$$

---

[3] We could try any real number greater than 4; as we will see momentarily, 6 has the advantage of leading to a sequence of natural numbers as the dimension increases.

[4] This is OEIS sequence A095264 [17].

LEMMA. $g_i = 2^{i+2} - 3i - 4$ satisfies the recurrence relation $g_{i+1} = 3g_i - 2g_{i-1} + 3$ for all $i \in \mathbb{Z}$.

*Proof.* This follows immediately from substituting the expressions for $g_i$ and $g_{i-1}$ into the right hand side of the recurrence relation:

$$3g_i - 2g_{i-1} + 3 = 3(2^{i+2} - 3i - 4) - 2(2^{i+1} - 3(i-1) - 4) + 3$$
$$= 2^{i+3} - 3i - 7 = 2^{i+3} - 3(i+1) - 4 = g_{i+1}. \quad \blacksquare$$

*Proof* (of Theorem 1). Consider the ultimate row in the Cayley-Menger determinant (3.3). When $n \geq 3$, if we subtract 3 times the previous row, add 2 times the row above that, and add 3 times the top row, the recurrence relation satisfied by $g_i$ (including the result that $g_{-1} = g_1 = 1$) and the fact that $1 = 3 \cdot 1 - 2 \cdot 1$ ensure that each of the off-diagonal elements in the bottom row will become 0. At the same time, the diagonal element will become $0 - 3(-g_1) + 2(-g_2) + 3 = 3 - 12 + 3 = -6$. As long as it is at least the fourth row, we can use the analogous linear combination of preceding rows to cancel all the below diagonal elements in the penultimate row, and again obtain $-6$ as the diagonal element. We can continue in this way until we get to the third row. Here we subtract the second row and add the first row to cancel the first two elements; this makes the diagonal element $g_1 + 1 = 2$. Finally, exchanging the top two rows gives an upper triangular matrix with diagonal elements $(1, 1, 2, -6, \ldots, -6)$, so the determinant of the matrix is $-2(-6)^{n-1}$. $\blacksquare$

The derivation of the spacetime Cayley-Menger determinant in the Appendix can be reversed: Since the determinant (3.3) is positive, when $n \leq d$ there exist points $p_0, \ldots, p_n \in \mathbb{M}^d$ with $p_i \prec p_j$ and spacetime separation $-s_{ij}^2 = -g_{j-i}$ when $i < j$. We name such a polytope, with these carefully chosen timelike edge lengths, a *timelike gem*. Figure 3 illustrates a 3-dimensional timelike gem, and the following theorem gives coordinates for the vertices of a timelike gem in every dimension.

THEOREM 2. *For $i \in \mathbb{N}$, let $p_i$ have coordinates*

$$\left(2^i - 1, (2^{i-1} - 1)\sqrt{3}, \ldots, (2^{i-(i-1)} - 1)\sqrt{3}\right) \quad (3.4)$$

*in $\mathbb{M}^i$ and identify $M^i \subset \mathbb{M}^d$ as the timelike subspace with coordinates $x_j = 0$ for*

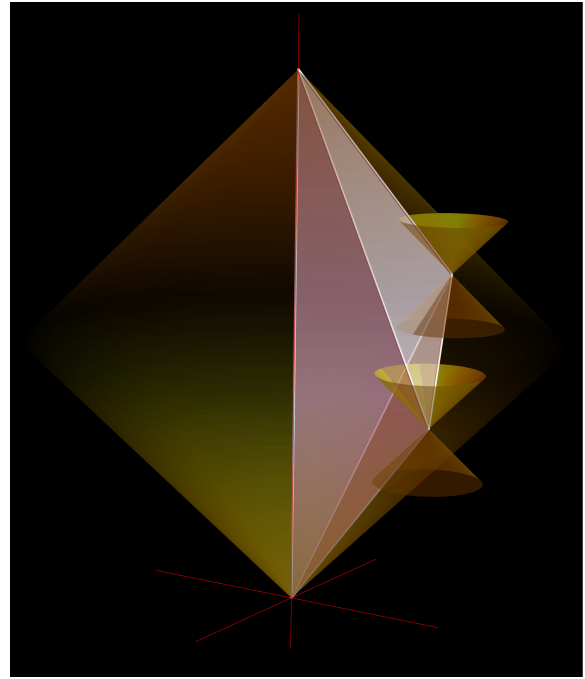

**Fig. 3.** A timelike gem in $\mathbb{M}^3$.

$i \leq j \leq d-1$. Then $p_0 p_1 \ldots p_n$ is an $n$-dimensional timelike gem, i.e., for all $0 \leq i < j \leq n$, $p_i \prec p_j$ and their spacetime separation is $-s_{ij}^2 = -g_{j-i}$.

*Proof.* Consider $i < j \in \mathbb{N}$. Then

$$-s_{ij}^2 = -(2^j - 2^i)^2 + 3 \sum_{k=1}^{i-1} (2^{j-k} - 2^{i-k})^2 + 3 \sum_{k=i}^{j-1} (2^{j-k} - 1)^2$$

$$= -(2^j - 2^i)^2 + 3(2^j - 2^i)^2 \sum_{k=1}^{i-1} 2^{-2k} + 3 \sum_{k=1}^{j-i} (2^k - 1)^2$$

$$= -(2^j - 2^i)^2 + (2^j - 2^i)^2 (1 - 2^{-2(i-1)}) + 4(2^{2(j-i)} - 1) + 12(1 - 2^{j-i}) + 3(j-i)$$

$$= -4 \cdot 2^{j-i} + 3(j-i) + 4 = -g_{j-i}. \quad \blacksquare$$

As we footnoted earlier, although this timelike gem with integer timelike separations is aesthetically pleasing, we need not choose $s_2^2 = 6$; it can be any real number greater than 4. The next theorem says that *any* such choice leads to sequences $s_i^2$ with positive spacetime volume.

THEOREM 3. *Suppose for $2 \leq n \in \mathbb{N}$ we have a sequence of positive numbers: $y_1, \ldots, y_n$, such that*

$$D_k = (-1)^k \det M_k = (-1)^k \det \begin{pmatrix} 0 & 1 & 1 & 1 & \cdots & 1 \\ 1 & 0 & -y_1 & -y_2 & \cdots & -y_k \\ 1 & -y_1 & 0 & -y_1 & \ddots & \vdots \\ 1 & -y_2 & -y_1 & 0 & \ddots & -y_2 \\ \vdots & \vdots & \ddots & \ddots & \ddots & -y_1 \\ 1 & -y_k & \cdots & -y_2 & -y_1 & 0 \end{pmatrix} > 0 \quad (3.5)$$

*for all $1 \leq k \leq n$. Then there exists $0 < y_{n+1} \in \mathbb{R}$ such that $D_{n+1} > 0$.*

*Proof.* As we remarked before Theorem 2, the derivation of the Cayley-Menger determinant of which (3.5) is a factor can be run in reverse to find a set of points $p_0 \prec \cdots \prec p_n \in \mathbb{M}^n$ such that for $i < j$, $\eta(p_j - p_i, p_j - p_i) = -y_{j-i}$, and the simplex with these vertices is nondegenerate since $D_n > 0$. The subsimplex $p_1 \prec \cdots \prec p_n$ is congruent to the subsimplex $p_0 \prec \cdots \prec p_{n-1}$, so there exists $q \in \mathbb{M}^n$ such that $p_n \prec q$ and $\eta(q, p_i) = \eta(p_n, p_{i-1})$ for all $1 \leq i \leq n$, i.e., the simplex $p_1 \prec \cdots \prec p_n \prec q$ is congruent to the simplex $p_0 \prec \cdots \prec p_n$.

Now identify this $\mathbb{M}^n \subset \mathbb{M}^{n+1}$ as the subspace defined by $x_n = 0$, and consider all the points $p \in \mathbb{M}^{n+1}$ such that $p_n \prec p$ and $\eta(p, p_i) = -y_{n+1-i}$, for all $1 \leq i \leq n$. Equivalently, define $F : \mathbb{M}^{n+1} \to \mathbb{R}^n$ by

$$F(p) = \begin{pmatrix} \eta(p - p_1, p - p_1) + y_n \\ \vdots \\ \eta(p - p_n, p - p_n) + y_1 \end{pmatrix},$$

and let $H = \{p_n \prec p \mid F(p) = 0\}$. $q \in H$ and the differential of $F$ at $q$ acting on $\Delta p \in \mathbb{M}^{n+1}$ is just

$$dF_q(\Delta p) = 2 \begin{pmatrix} \eta(\Delta p, q - p_1) \\ \vdots \\ \eta(\Delta p, q - p_n) \end{pmatrix}.$$

Since $\{q - p_i \mid 1 \leq i \leq n\}$ are linearly independent (because the simplex $p_1 \prec \cdots \prec p_n \prec q$ is nondegenerate), $dF_q(\Delta p) = 0$ only for $\Delta p \propto \hat{x}_n$. That is, $H$ is one-dimensional and transverse to $\mathbb{M}^n$. Thus any open neighborhood of $q$ intersects $H$ in a point $p_{n+1} \notin \mathbb{M}^n$. By construction $\eta(p_{n+1}-p_i, p_{n+1}-p_i) = -y_{n+1-i}$ for $1 \leq i \leq n$, and $\eta(p_{n+1}-p_0, p_{n+1}-p_0) < 0$ since $p_0 \prec p_1 \prec \cdots \prec p_n \prec p_{n+1}$. Set $y_{n+1} = -\eta(p_{n+1} - p_0, p_{n+1} - p_0) > 0$, and observe that since $p_{n+1} \notin \mathbb{M}^n$, $D_{n+1} > 0$. ∎

We remark that there are, consequently, many incongruent timelike gems, facetted differently. For example, if we choose $s_2^2 = 5$ and subsequently always choose the smallest possible integer value for $s_i^2$, we obtain, charmingly, the *pentagonal numbers*, $f_n$, which satisfy the recurrence relation $f_{n+1} = 2f_n - f_{n-1} + 3$, with $f_0 = 0$ and $f_1 = 1$.[5] This recurrence relation enables a proof—that this sequence of timelike separations makes the spacetime volumes positive—entirely analogous to the proof of Theorem 1.

## 4. Computing spacelike separations in any dimension

Having established the existence of timelike gems with Theorems 1, 2, and 3, we can use them to compute the distance between spacelike related points $a, b \in \mathbb{M}^d$, for any $0 < d \in \mathbb{N}$. While we cannot directly locate all the points in the hyperplane $M$ spanned by $H_{ab}(s, s)$, varying $0 < s \in \mathbb{R}$ foliates the subset of $M$ that is timelike related to both $a$ and $b$ with hyperbolae. Let $E_{ab}^-$ denote the component of this subset to the past of $a$ and $b$, and $E_{ab}^+$ the component to the future. We begin by choosing an arbitrary point, $c_0 \in E_{ab}^+$.[6] Next, we choose points sequentially with

$$c_j \in E_{ab}^+ \cap \bigcap_{i=0}^{j-1} H_{c_i}^+(-g_{j-i}). \tag{4.1}$$

$c_0 c_1 \ldots c_j$ is a timelike gem spanning a $j$-dimensional timelike subspace, *and is unique up to Poincaré transformations*, by construction. Thus, the largest $j \in \mathbb{N}$ for which the intersection (4.1) is nonempty is the dimension of $M$, i.e., $d - 1$, and $c_0 c_1 \ldots c_{d-1}$ is a timelike gem with spacetime volume $B = \sqrt{3^{d-2}}/(d-1)!$. Taking the timelike gem as the base of the congruent $d$-simplices with $a$ or $b$ as the apex, we have

$$\left(\mathsf{stvol}(c_0 c_1 \ldots c_{d-1} a)\right)^2 = \left(\frac{Bh}{d}\right)^2 = \frac{3^{d-2} h^2}{(d!)^2}, \tag{4.2}$$

---

[5] This is OEIS sequence A000326 [17].

[6] We note that Boguñá and Krioukov choose an element (approximately) in $E_{ab}^-$ to regulate their estimation of the spacelike distance between unrelated elements in a causal set [11].

where $(2h)^2 = \eta(b-a, b-a)$ as in §2. Now let $-s_i^2 = \eta(c_i - a, c_i - a)$. Then we also have

$$\left(\mathsf{stvol}(c_0 c_1 \ldots c_{d-1} a)\right)^2 = \frac{(-1)^d}{2^d (d!)^2} \det \begin{pmatrix} 0 & 1 & 1 & \cdots & 1 & 1 \\ 1 & 0 & -g_1 & \cdots & -g_{d-1} & -s_0^2 \\ 1 & -g_1 & 0 & \ddots & \vdots & -s_1^2 \\ \vdots & \vdots & \ddots & \ddots & -g_1 & \vdots \\ 1 & -g_{d-1} & \cdots & -g_1 & 0 & -s_{d-1}^2 \\ 1 & -s_0^2 & -s_1^2 & \cdots & -s_{d-1}^2 & 0 \end{pmatrix}. \quad (4.3)$$

Combining (4.2) and (4.3) gives the spacelike separation of $a$ and $b$:

$$\eta(b-a, b-a) = (2h)^2 = \frac{(-1)^d}{6^{d-2}} \det \begin{pmatrix} 0 & 1 & 1 & \cdots & 1 & 1 \\ 1 & 0 & -g_1 & \cdots & -g_{d-1} & -s_0^2 \\ 1 & -g_1 & 0 & \ddots & \vdots & -s_1^2 \\ \vdots & \vdots & \ddots & \ddots & -g_1 & \vdots \\ 1 & -g_{d-1} & \cdots & -g_1 & 0 & -s_{d-1}^2 \\ 1 & -s_0^2 & -s_1^2 & \cdots & -s_{d-1}^2 & 0 \end{pmatrix}.$$

To reconstruct $\mathbb{M}^d$ as a vector space, pick any point $p_0$, construct a timelike gem $p_0 p_1 \ldots p_d$, and assign coordinates to its vertices as in Theorem 2. Since we know timelike separations and we have just shown that we can compute spacelike separations, the spacetime separations $\ell_i$ of any other point $q \in \mathbb{M}^d$ from $p_0, \ldots, p_d$ specify $d+1$ quadratic equations:

$$\eta(q - p_i, q - p_i) = \ell_i.$$

Subtracting the $p_0$ equation from each of the others gives $d$ *linear* equations:

$$-2\eta(q, p_i - p_0) = \ell_i - \ell_0 + g_i, \quad (4.4)$$

since $\eta(p_0, p_0) = 0$ and $\eta(p_i, p_i) = -g_i$. Because the timelike gem has positive volume, it is not degenerate, so the vectors $p_i - p_0$ span $\mathbb{M}^d$; thus (4.4) has a unique solution for the coordinates of $q$. Using this coordinate system we can define vector addition and scalar multiplication, proving our main theorem:

THEOREM 4. *For $0 < d \in \mathbb{N}$, given the set of points in $\mathbb{M}^d$, together with their causal relations and the spacetime separations of pairs of causally related points, we can reconstruct the vector space structure of the set, and in particular, the spacetime separations of pairs of spatially related points.*

## 5. Conclusions

We have shown that for Minkowski space of any dimension, given the causal relations among its points and the spacetime separations of causally related points, we can compute the spacetime separations of spatially related points and also reconstruct its vector space structure. An analogous approach also proves that given the causal structure and the spacetime separations of *spatially* related points we can compute the spacetime separations of causally related points: Given two such points $p \prec q$, consider the intersection of the *timelike* hyperboloids consisting of points a fixed spacelike distance from $p$ or $q$, respectively. The intersection is a sphere, from which we can select $1 < k \in \mathbb{N}$ points. Since these points are spacelike related, their spacetime separations are given. The largest $k$ for which the Euclidean Cayley-Menger determinant is nonzero is the dimension of the Minkowski space, whence these $k$ points form the base of congruent simplices with additional vertex $p$ or $q$; computing the spacetime volume of either of these determines (half) the timelike distance between $p$ and $q$, just as it did for (half) the spacelike distance between $a$ and $b$ in §4.

While this last observation is of geometrical interest, it has little relevance to the causal set program for quantum gravity that motivates our main result. As we explained in the Introduction, we have investigated the consequence of knowing spacetime separations of causally related points because these can be estimated in causal sets arising from a Poisson process in Minkowski space. The fact that these estimates converge to the true separations in the limit of infinitely dense Poisson processes [12,13] shows that the spacetime separations, computed as we did in §4, also converge to the true values. This raises the same question of bounding the expected error in the approximations of spacetime separations and the vector space structure obtained from a finite density Poisson process that Boguñá and Krioukov raise in their work [11]. Conceivably the rate of convergence may differ for different estimation algorithms; we leave this analysis for the future.

Finally, since the theorem of Hawking [6] and Malament [7] is the general spacetime analogue of the theorem of Aleksandrov [8,9] and Zeeman [10] for Minkowski space, we are optimistic that the ideas we have used here to prove that spatial separations can be computed from causal separations in Minkowski space may be generalized to curved spacetimes. This would be a substantial step towards proving the *Hauptvermutung* of causal set theory.

**Acknowledgements**

We thank Orest Bucicovschi for pointing us towards Aleksandrov's work, Ryshard-Pavel Kostecki for archiving a copy of [9], Alexander Guts for describing the extent and context of Aleksandrov's program [23] and providing Ovchinnikova's given name, and a curious referee for asking the question that motivated Theorem 3.

**Appendix: The Cayley-Menger determinant for spacetime volume**

Recall that the (spacetime) volume of a parallelotope defined by a linearly independent set of vectors $\{v_j\}$ in either Euclidean or Minkowski space is the absolute value of the determinant of the matrix with entries $V_{ij} = (v_j)_i$. Furthermore, that the square of the spacetime volume of a simplex in Minkowski space is given by a Cayley-Menger determinant [14,15] has been noted multiple times [18,19,20,21], initially in the context of the Regge calculus [22]. We provide an elementary derivation here so that we can refer to a consequence of it in §3.

An $n$-simplex in Minkowski space is specified by its vertices $p_i$, $i \in \{0, \ldots, n\}$. Just as in Euclidean space, its spacetime volume is determined by the $n+1$ vertex vectors, $p_i$. In coordinates this is exemplified by the formula for the (spacetime) volume of a tetrahedron,

$$\text{stvol}(\text{tetrahedron}) = \frac{1}{3!} \det \begin{pmatrix} 1 & 1 & 1 & 1 \\ t_0 & t_1 & t_2 & t_3 \\ x_0 & x_1 & x_2 & x_3 \\ y_0 & y_1 & y_2 & y_3 \end{pmatrix}, \tag{A.1}$$

which we recognize as the determinant of the spanning vectors, $p_i - p_0$, obtained by subtracting the first column from each of the others. The division by 3!, or more generally, $n!$, accounts for the fact that an $n$-cube can be partitioned into $n$ pyramids with $(n-1)$-cube bases, and then those $(n-1)$-cubes can be partitioned, recursively—and that these combinatorics are invariant under linear transformations so they apply as well to the parallelotope spanned by these vectors. Our goal is to rewrite the determinant (A.1) in terms of the spacetime separations of the vertices of the $n$-simplex, just as the Cayley-Menger determinant does in Euclidean space [14,15].

We begin by increasing the size of the matrix, keeping the value of the determinant the same, and continue to illustrate the derivation with a tetrahedron ($n = 3$):

$$\text{stvol}(\text{tetrahedron}) = \frac{1}{3!} \det \begin{pmatrix} 1 & \eta(p_0, p_0) & \eta(p_1, p_1) & \eta(p_2, p_2) & \eta(p_3, p_3) \\ 0 & 1 & 1 & 1 & 1 \\ 0 & t_0 & t_1 & t_2 & t_3 \\ 0 & x_0 & x_1 & x_2 & x_3 \\ 0 & y_0 & y_1 & y_2 & y_3 \end{pmatrix}, \tag{A.2}$$

where the top row entries are the Minkowski scalar products. We call this matrix $U$, and define $W$ to be the $(n+2) \times (n+2)$ matrix:

$$W = \begin{pmatrix} 0 & 1 & 0 & 0 & 0 \\ 1 & 0 & 0 & 0 & 0 \\ 0 & 0 & 2 & 0 & 0 \\ 0 & 0 & 0 & -2 & 0 \\ 0 & 0 & 0 & 0 & -2 \end{pmatrix}, \tag{A.3}$$

with $\det W = (-1)^n 2^n$, since the first 2 is positive. Now,

$$U^\mathsf{T} W U$$
$$= \begin{pmatrix} 1 & 0 & 0 & 0 & 0 \\ \eta(p_0, p_0) & 1 & t_0 & x_0 & y_0 \\ \eta(p_1, p_1) & 1 & t_1 & x_1 & y_1 \\ \eta(p_2, p_2) & 1 & t_2 & x_2 & y_2 \\ \eta(p_3, p_3) & 1 & t_3 & x_3 & y_3 \end{pmatrix} \begin{pmatrix} 0 & 1 & 1 & 1 & 1 \\ 1 & \eta(p_0, p_0) & \eta(p_1, p_1) & \eta(p_2, p_2) & \eta(p_3, p_3) \\ 0 & 2t_0 & 2t_1 & 2t_2 & 2t_3 \\ 0 & -2x_0 & -2x_1 & -2x_2 & -2x_3 \\ 0 & -2y_0 & -2y_1 & -2y_2 & -2y_3 \end{pmatrix},$$

which multiplies out to

$$\begin{pmatrix} 0 & 1 & 1 & 1 & 1 \\ 1 & 0 & \eta(p_0-p_1,p_0-p_1) & \eta(p_0-p_2,p_0-p_2) & \eta(p_0-p_3,p_0-p_3) \\ 1 & \eta(p_1-p_0,p_1-p_0) & 0 & \eta(p_1-p_2,p_1-p_2) & \eta(p_1-p_3,p_1-p_3) \\ 1 & \eta(p_2-p_0,p_2-p_0) & \eta(p_2-p_1,p_2-p_1) & 0 & \eta(p_2-p_3,p_2-p_3) \\ 1 & \eta(p_3-p_0,p_3-p_0) & \eta(p_3-p_1,p_3-p_1) & \eta(p_3-p_2,p_3-p_2) & 0 \end{pmatrix},$$

since $\eta(p_i,p_i) + 2t_it_j - 2x_ix_j - 2y_iy_j + \eta(p_j,p_j) = \eta(p_i,p_i) - 2\eta(p_i,p_j) + \eta(p_j,p_j) = \eta(p_i-p_j, p_i-p_j)$. Let $\ell_{ij}$ denote the spacetime separation between points $i$ and $j$. Thus we have:

$$\bigl(\text{stvol}(n\text{-simplex})\bigr)^2 = \frac{(-1)^n}{2^n(n!)^2} \det \begin{pmatrix} 0 & 1 & 1 & 1 & \cdots & 1 \\ 1 & 0 & \ell_{01} & \ell_{02} & \cdots & \ell_{0n} \\ 1 & \ell_{10} & 0 & \ell_{12} & \cdots & \ell_{1n} \\ 1 & \ell_{20} & \ell_{21} & 0 & \cdots & \ell_{2n} \\ \vdots & \vdots & \vdots & \vdots & \ddots & \vdots \\ 1 & \ell_{n0} & \ell_{n1} & \ell_{n2} & \cdots & 0 \end{pmatrix}.$$

This is the Cayley-Menger determinant for the square of spacetime volume. As in Euclidean space, an $n$-simplex with specified spacetime separations between its vertices is only realizable by $n+1$ points in Minkowski space if this determinant (including the prefactors) is positive (and all the subsimplices are also realizable).

# References

[1] Luca Bombelli, Joohan Lee, David A. Meyer and Rafael D. Sorkin, "Space-time as a causal set", *Physical Review Letters* **59** (1987) 521–524.

[2] Rafael D. Sorkin, "Spacetime and causal sets", in Juan Carlos D'Olivo, Eduardo Nahmad-Achar, Michael P. Ryan, Jr., Louis F. Urrutia and Federico Zertuche, eds., *Relativity and Gravitation: Classical and Quantum*, proceedings of the 7th Latin American Symposium on Relativity and Gravitation (SILARG VII), Cocoyoc, Mexico, 2–8 December 1990 (Singapore: World Scientific 1991) 150–173.

[3] John W. Milnor, "Two complexes which are homeomorphic but combinatorially distinct", *Annals of Mathematics* **74** (1961) 575–590.

[4] Ernst Steinitz, "*Beiträge zur Analysis situs*", *Sitzungsberichte der Berliner Mathematischen Gesellschaft* **7** (1908) 29–49.

[5] Heinrich Tietze, "*Über die topologischen Invarianten mehrdimensionaler Mannigfaltigkeiten*", *Monatshefte für Mathematik und Physik* **19** (1908) 1–118.

[6] Stephen W. Hawking, Andrew R. King and Patrick J. McCarthy, "A new topology for curved space-time which incorporates the causal, differential, and conformal structures", *Journal of Mathematical Physics* **17** (1976) 174–181.

[7] David B. Malament, "The class of continuous timelike curves determines the topology of spacetime", *Journal of Mathematical Physics* **18** (1977) 1399–1404.

[8] Александр Д. Александров, "О преобразованиях Лоренца", *Успехи Математических Наук* **5** (1950) 187.

[9] Александр Д. Александров и Валентина В. Овчинникова, "Замечания к основам теории относительности", *Вестник Ленинградского университета. Серия математики, механики и астрономии* (1953) No. 11, 95–110.

[10] E. Christopher Zeeman, "Causality implies the Lorentz group", *Journal of Mathematical Physics* **5** (1964) 490–493.

[11] Marián Boguñá and Dmitri Krioukov, "Measuring spatial distances in causal sets via causal overlaps", *Physical Review D* **110** (2024) 024008/1–12.

[12] Graham Brightwell and Ruth Gregory, "Structure of random discrete spacetime", *Physical Review Letters* **66** (1991) 260–263.

[13] Béla Bollobás and Graham Brightwell, "Box spaces and random partial orders", *Transactions of the American Mathematical Society* **324** (1991) 59–72.

[14] Arthur Cayley, "On a theorem in the geometry of position", *Cambridge Mathematical Journal* **II** (1841) 267–271.

[15] Karl Menger, "*Untersuchungen über allgeneine Metrik*", *Mathematische Annalen* **100** (1928) 75–163.

[16] 華嚴五教止觀 (*Huáyán wǔjiào zhǐguān*), CBETA Chinese Electronic Tripitaka Collection T45n1867; translated as *Cessation and Contemplation in the Five Teachings of Hua-yen* in Thomas Cleary, *Entry into the Inconceivable: An Introduction to Hua-yen Buddhism* (Honolulu: University of Hawaii Press 1983).

[17] *The On-Line Encyclopedia of Integer Sequences*, https://oeis.org (2010) Sequence A095264 and Sequence A000326.

[18] John Archibald Wheeler, "Geometrodynamics and the issue of the final state", in Cécile DeWitt and Bryce DeWitt, eds., *Relativité, Groupes et Topologie : Relativity, Groups and Topology*, lectures delivered at Les Houches during the 1963 session of the Summer School of Theoretical Physics, University of Grenoble (New York: Gordon and Breach 1964) 315–520.

[19] Rafael D. Sorkin, *The Development of Simplectic Methods for the Metrical and Electromagnetic Fields*, Physics Ph. D. Thesis (California Institute of Technology 1974).

[20] Herbert W. Hamber, "Quantum gravity on the lattice", *General Relativity and Gravitation* **41** (2009) 817–876.

[21] Kyle Tate and Matt Visser, "Realizability of the Lorentzian $(n,1)$-simplex", *Journal of High Energy Physics* **2012** (2012) article number 28/1–13.

[22] Tullio Regge, "General relativity without coordinates", *Il Nouvo Cimento* **XIX** (1961) 558–571.

[23] Александр К. Гуц, *Хроногеометрия: Аксиоматическая теория относительности* (Омск: УниПак 2008).